\documentclass[sigconf]{acmart}
\usepackage{multirow} 
\usepackage{hhline} 
\usepackage{subcaption}
\usepackage{enumitem}
\usepackage{array} 
\usepackage{algorithm}
\usepackage{algpseudocode}
\usepackage{balance}

\AtBeginDocument{%
  }

\newcommand{\bmm}{\mathbf{m}}

\newcommand{\bx}{\mathbf{x}}

\newcommand{\by}{\mathbf{y}}

\copyrightyear{2026}
\acmYear{2026}
\setcopyright{cc}
\setcctype{by}
\acmConference[ICTIR '26]{Proceedings of the 2026 International ACM SIGIR Conference on Innovative Concepts and Theories in Information Retrieval (ICTIR)}{July 25, 2026}{Melbourne, VIC, Australia}
\acmBooktitle{Proceedings of the 2026 International ACM SIGIR Conference on Innovative Concepts and Theories in Information Retrieval (ICTIR) (ICTIR '26), July 25, 2026, Melbourne, VIC, Australia}
\acmDOI{10.1145/3805713.3820399}
\acmISBN{979-8-4007-2600-2/2026/07}

\begin{document}
\sloppy
\title{RAMP: Robust Ad Recommendation Under Limited Personalized-Feature Availability via Masking and Alignment Pathways}
\author{Dairui Liu}
\authornote{Both authors contributed equally to this research.}
\email{dairui.liu@ucd.ie}
\author{Zhongyi Lu}
\authornotemark[1]
\email{zhongyi.lu@ucd.ie}
\affiliation{
  \institution{University College Dublin}
  \city{Dublin}
  \country{Ireland}
}

\author{Roger Zhe Li}
\email{roger.zhe.li@huawei.com}
\affiliation{%
  \institution{Huawei Ireland Research Center}
  \city{Dublin}
  \country{Ireland}}

\author{Changhong Jin}
\author{Jitao Lu}
\email{changhong.jin@ucd.ie}
\email{jitao.lu@ucd.ie}
\affiliation{%
  \institution{University College Dublin}
  \city{Dublin}
  \country{Ireland}
}

\author{Xinyang Shao}
\email{xinyang.shao@huawei-partners.com}
\affiliation{%
  \institution{Huawei Ireland Research Center}
  \city{Dublin}
  \country{Ireland}}

\author{Bichen Shi}
\email{bichen.shi@huawei-partners.com}
\affiliation{%
  \institution{Huawei Ireland Research Center}
  \city{Dublin}
  \country{Ireland}}

\author{Mete Sertkan}
\email{mete.sertkan@h-partners.com}
\affiliation{%
  \institution{Huawei Ireland Research Center}
  \city{Dublin}
  \country{Ireland}}

\author{Aghiles Salah}
\email{aghiles.salah@h-partners.com}
\affiliation{%
  \institution{Huawei Ireland Research Center}
  \city{Dublin}
  \country{Ireland}}

\author{Aonghus Lawlor}
\email{aonghus.lawlor@ucd.ie}
\affiliation{%
 \institution{University College Dublin}
  \city{Dublin}
  \country{Ireland}}

\author{Barry Smyth}
\email{barry.smyth@ucd.ie}
\affiliation{%
 \institution{University College Dublin}
  \city{Dublin}
  \country{Ireland}}

\author{Tri Kurniawan Wijaya}
\email{tri.kurniawan.wijaya@huawei.com}
\affiliation{%
  \institution{Huawei Ireland Research Center}
  \city{Dublin}
  \country{Ireland}}

\author{Ruihai Dong}
\email{ruihai.dong@ucd.ie}
\affiliation{%
 \institution{University College Dublin}
  \city{Dublin}
  \country{Ireland}}

\author{Xingsheng Guo}
\email{xingsheng.guo1@huawei-partners.com}
\affiliation{%
  \institution{Huawei Ireland Research Center}
  \city{Dublin}
  \country{Ireland}}


\renewcommand{\shortauthors}{Dairui Liu et al.}

\begin{abstract}
  Click-through rate (CTR) and conversion rate (CVR) prediction are fundamental tasks in online advertising, aiming to estimate the likelihood of user interactions based on various features. 
  While personalized attributes such as age and gender can significantly enhance predictive accuracy, their use is increasingly restricted by privacy regulations, thereby limiting available data for both training and inference. To address this challenge, we propose RAMP (\textbf{R}obust \textbf{A}d Recommendation Under Limited Personalized-Feature Availability via \textbf{M}asking and Alignment \textbf{P}athways), which is designed to improve CTR/CVR prediction accuracy when personalized features are not accessible, thus supporting deployment in privacy-constrained settings.
  RAMP consists of (i) a personalized pathway built upon a dual-tower component with identical inputs but independent parameters, where output masking separates predictions for personalized and non-personalized signals, (ii) a separate non-personalized pathway trained with non-personalized features only, and (iii) a distillation-inspired prediction-alignment architecture between (i) and (ii) that improves prediction when personalized features are unavailable.
  We conduct comprehensive experiments using both public benchmarks and industrial datasets to evaluate the performance of RAMP. Our evaluation spans multiple backbone models and different settings: with and without access to personalized features. The results show that RAMP consistently outperforms state-of-the-art methods when personalized features are missing, while maintaining competitive performance when all features are available. 
  Our code is publicly available at \url{https://github.com/Ruixinhua/RAMP}.
\end{abstract}




\maketitle

\section{Introduction}\label{sec:intro}
Online advertising relies on predicting user responses, such as click-through rate (CTR) and conversion rate (CVR), to guide ad delivery and optimize performance~\cite{Zhu_Jin_Tan_Pan_Zeng_Li_Gai_2017,Yang_Pan_2022}. These predictions estimate the likelihood of user interactions along the typical behavioral sequence of impression $\rightarrow$ click $\rightarrow$ conversion~\cite{Ma_Zhao_Huang_Wang_Hu_Zhu_Gai_2018}, and play a central role in improving both user experience and business outcomes~\cite{Li_Zhang_Zhang_Li_Sang_Zhu_2024,Chang_Zhang_Fu_Zang_Guan_Lu_Hui_Leng_Niu_Song_Gai_2023,Chen_Wang_Liu_Tang_Guo_Zheng_Yao_Zhang_He_2021}.
However, in practice, these models often rely on personalized user features, which may be partially or entirely unavailable due to user consent constraints. This results in a feature-missing scenario that can significantly degrade prediction accuracy.

\begin{figure}[t]
    \centering
    \includegraphics[width=0.8\linewidth]{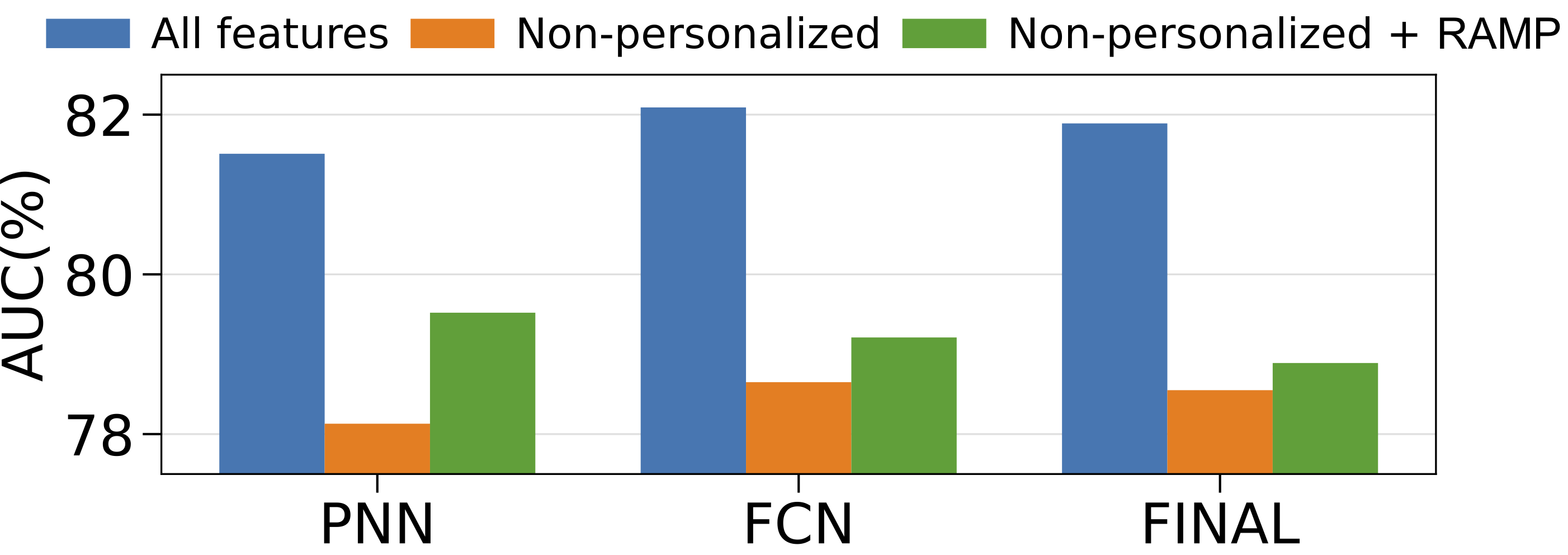}
    \Description{A bar chart showing AUC performance on the CriteoPrivateAd dataset. Strong baselines like PNN, FCN, and FINAL show a significant drop in AUC when personalized features are removed. The proposed RAMP method, when applied to these backbones, recovers a substantial portion of this lost accuracy, demonstrating improved robustness under feature constraints.}
    \caption{\small Motivation study based on the CriteoPrivateAd dataset. Removing personalized features causes a substantial performance drop for strong CTR/CVR baselines (PNN~\cite{DBLP:conf/icdm/QuCRZYWW16}, FCN~\cite{Li_Zhang_Zhang_Li_Sang_Zhu_2024}, FINAL~\cite{FinalMLP}). Applying RAMP on top of each backbone partially recovers the lost accuracy, narrowing the gap under limited personalized features.}
\label{fig:motivation_criteo_feature_constraints}
\end{figure}

Deep neural networks have become the \textit{de facto} choice for CTR/CVR prediction because they capture high-order feature interactions and scale to industrial data volumes~\cite{DBLP:conf/ijcai/ZhangQGTH21, Li_Zhang_Zhang_Li_Sang_Zhu_2024}.
Their accuracy, however, often hinges on leveraging rich user- (e.g., \texttt{gender}), context- (e.g., \texttt{site\_id}), and ad-side (e.g., \texttt{campaign\_id}) features; especially signals tied to user identity. 
Integrating such sensitive information via complex, opaque interactions raises privacy and compliance concerns~\cite{DBLP:journals/csur/LiuDSRFL21} under regulations such as the GDPR and China’s Personal Information Protection Law.

\textbf{Personalized vs.\ non-personalized.} We distinguish features by whether they are \emph{user-linked} and thus subject to consent constraints. \emph{Personalized features} include user IDs, device/account attributes, and behavioral history, while \emph{non-personalized features} include context (e.g., query/page), coarse time/location, and item attributes. In practice, systems face \emph{feature-availability constraints}, where personalized features may be missing during training or inference (e.g., non-consented traffic), reducing feature coverage and potentially degrading prediction performance (Figure~\ref{fig:motivation_criteo_feature_constraints}). We call traffic \emph{personalized} when both feature types are available, and \emph{non-personalized} when only non-personalized features are used. In our offline experiments on public datasets, non-personalized traffic is approximated by masking personalized features, which serves as a practical simulation of feature unavailability under consent constraints, although it may not fully capture all characteristics of real non-consented traffic.

Despite these concerns, relatively little work has explicitly focused on improving CTR/CVR prediction when only non-personalized features are available. Existing privacy-oriented approaches primarily focus on ensuring data privacy or compliance, rather than maintaining prediction performance under feature-availability constraints. For example, system-level paradigms such as federated learning~\cite{Han_Ma_Mei_Liu_2021, Deng_Wang_Yue_Rao_Zu_Wang_Chen_Ren_Zhang_2024, DBLP:journals/apin/MaZZDW24} enable collaborative training without centralized data sharing, but focus on privacy preservation and introduce additional system complexity in practice. Other approaches target specific application scenarios~\cite{Rafieian_Yoganarasimhan_2020}, which may limit their generalizability across platforms and feature spaces. In parallel, preprocessing-based data augmentation methods, such as injecting structured knowledge via knowledge graphs~\cite{DBLP:journals/tomccap/LiuSSNWL24, DBLP:journals/apin/MaZZDW24}, can partially compensate for missing user signals, but typically require substantial curation and depend on the coverage and quality of external knowledge. Importantly, our goal is not to provide formal privacy guarantees, but to maintain robust prediction performance when consent-dependent personalized features are unavailable.

\textbf{Why a dual-tower design for personalized pathway.} This gap motivates an \textit{in-processing} approach that improves learning under feature-availability constraints without system changes. The key challenge is the mixed regimes---personalized traffic with full inputs and non-personalized traffic with restricted inputs---so the model effectively faces two related but different prediction problems. A single network trained across both regimes (e.g., by feature masking) can suffer from negative transfer, as gradients originating from personalized and non-personalized traffic may optimize representations toward different objectives. Consequently, the shared model tends to overfit interactions that depend on personalized signals, leading to degraded performance when such signals are absent.



We address this with a \emph{personalized pathway} containing two towers. Both towers share the same embedding layer and training feature set but use independent prediction networks. Leveraging output masking based on traffic's personalization preference, we maintain a tower dedicated to personalized prediction while the other is dedicated to non-personalized prediction.

This separation reduces cross-regime interference but does not fully close the performance gap on non-personalized traffic, where the model operates under a reduced feature set. To directly address this setting, we introduce an additional \emph{non-personalized} pathway, which is trained using only non-personalized features. This separate pathway allows the model to learn knowledge when a controlled network where only non-personalized features are accessible and acts as an enhancement to the dual-tower component. 

We apply a lightweight loss to align the personalized and the non-personalized pathways, enabling the knowledge from richer signals while remaining fully usable without personalized features. This mechanism shares many similarities to the concept of knowledge distillation, which allows the behavior of two models to be matching. Therefore, we name this as a distillation-inspired prediction-alignment architecture. In the proposed architecture, both pathways use the same input, which contains personalized and non-personalized traffic. The non-personalized pathway is used while training but is not required for online inference. 

Building on these ideas, we propose a model-agnostic \textbf{R}obust \textbf{A}d recommendation under limited Personalized-Feature availability via \textbf{M}asking and alignment \textbf{P}athways  (RAMP) architecture for CTR/CVR prediction under personalized-feature constraints. RAMP combines (i) a personalized pathway built upon a personalization-gated dual-tower foundation that separates supervision for personalized and non-personalized traffic, (ii) a non-personalized pathway trained with non-personalized features only, and (iii) a knowledge-distillation-inspired architecture between (i) and (ii) that enables accurate prediction when personalized features are constrained. Experiments on public benchmarks and a large-scale industrial dataset show that RAMP improves non-personalized performance for both CTR and CVR prediction while remaining competitive when personalized signals are available.

The contributions of our work are summarized as follows:


\begin{itemize}
   \item We propose a personalized pathway built upon a dual-tower component trained on the same data. Using selective masking conditioned on user's personalization preference, one tower specializes in personalized prediction while the other focuses on non-personalized prediction; 
   \item We introduce a non-personalized pathway that is only trained with non-personalized features, enhancing the dual-tower component when only non-personalized features are available.
   \item We introduce a distillation-inspired prediction alignment architecture between the personalized and the non-personalized pathway. By minimizing the discrepancy between their predictions, the behavior of the two pathways converges, which improves prediction accuracy on samples with only non-personalized features.
   \item We conduct extensive experiments on widely used benchmark datasets and a large-scale industrial dataset (covering both CTR and CVR), and further evaluate non-personalized recommendation scenarios to demonstrate the effectiveness of RAMP for privacy-aware advertising prediction.
\end{itemize}

\section{Related Work}
\label{sec:related_work}

\subsection{Deep Learning-based CTR/CVR Prediction}
\label{sec:related_work:deep_ctr_model}

CTR/CVR prediction is a closely related task in computational advertising. Although CVR involves task-specific challenges (e.g., delayed feedback and data sparsity), it shares the same core setup as CTR prediction, and many CTR models can be adapted to CVR prediction with minor modifications.

Deep models are now the dominant paradigm for CTR/CVR prediction due to their ability to represent heterogeneous features and capture high-order interactions~\cite{DBLP:conf/ijcai/ZhangQGTH21}. Existing methods can be broadly grouped into: (i) \emph{interaction modeling}, which explicitly parameterizes high-order feature crosses (e.g., Wide \& Deep~\cite{wide_deep_2016}, DeepFM~\cite{Guo_Tang_Ye_Li_He_2017}, xDeepFM~\cite{Lian_Zhou_Zhang_Chen_Xie_Sun_2018}, DCN/DCNv2~\cite{Wang_Fu_Fu_Wang_2017, dcnv2}, AutoInt~\cite{song2019autoint}); (ii) \emph{feature importance/selection}, which learns to reweight or select informative fields (e.g., FiBiNet~\cite{fibinet}, MaskNet~\cite{Wang_She_Zhang_2021}, FCN~\cite{Li_Zhang_Zhang_Li_Sang_Zhu_2024}, FinalMLP~\cite{FinalMLP}); and (iii) \emph{representation enhancement}, which improves embeddings or feature encoders for sparse categorical inputs~\cite{DBLP:conf/sigir/WangWLGL0G22, DBLP:conf/cikm/Cheng22, DBLP:conf/cikm/HeCWJYX21}. 
In addition, multi-task and multi-scenario learning has become increasingly popular for CTR/CVR prediction to share information across related labels or traffic segments~\cite{DBLP:conf/recsys/LiuLSLSWWGSLLLJ25, DBLP:conf/recsys/DishiFKSS25, DBLP:conf/www/LiuJWLDB0T23}.


These methods perform well with rich user-linked signals, but prioritize full feature utilization and do not explicitly address \emph{feature availability constraints}, leading to degraded performance when personalized features are missing for part of the traffic.

In recent years, knowledge distillation has been widely adopted in recommender systems to transfer knowledge from complex teacher models to efficient students under practical constraints~\cite{DBLP:conf/cikm/KangHKY20}. It can be broadly categorized into several paradigms: prediction-level distillation for soft-target transfer, as introduced in the seminal work by Geoffrey Hinton et al.~\cite{hinton2015distilling} and extended in recommendation scenarios~\cite{DBLP:conf/www/LiangL0LSZZCZL025}; representation-level distillation for aligning latent embeddings~\cite{DBLP:conf/cikm/KangHKY20}, including recent advances that leverage latent alignment under privileged information settings~\cite{DBLP:conf/kdd/Yuan0HD25}; structure-level distillation for preserving user–item or item–item relations~\cite{DBLP:journals/jiis/MaXL25}; bias-aware or group-wise distillation to address popularity bias~\cite{DBLP:conf/recsys/LiuLZTZPLS22}; and data- or teacher-enhanced distillation with external supervision signals, such as those incorporating privileged features in large-scale industrial systems~\cite{DBLP:conf/kdd/XuLGGYPSWSO20,DBLP:conf/aaai/Du0WCZO25}.

Our method shares commonalities with prediction-level knowledge distillation but does not fall into the standard teacher-student structure. Instead, our two pathways, the personalized and the non-personalized pathways, align knowledge with each other to enhance CTR/CVR prediction when personalized features are constrained.
\subsection{Non-Personalized Recommendation}
\label{sec:related_work:non_per_recomm}
Recommender systems have long relied on personalization to match users with content~\cite{recsys_survey_ieee}. In practice, however, privacy regulations and consent constraints increasingly limit the collection and use of user-linked signals~\cite{DBLP:conf/ecir/Mullner23}, yielding traffic where only non-personalized context and item/ad attributes are available. This ``non-personalized traffic'' creates a distinct accuracy challenge in advertising prediction. 

Prior work related to non-personalized recommendation often focuses on bias mitigation~\cite{DBLP:conf/cikm/JingZ0023} or beyond-accuracy objectives~\cite{DBLP:journals/tweb/RanaSBCF24}. Comparatively less attention has been paid to \emph{accuracy-oriented} modeling that explicitly bridges the gap between personalized and non-personalized feature regimes in computational advertising. Our work complements existing privacy-preserving directions by focusing on model-side learning objectives and structured knowledge transfer under feature-availability constraints. We note that while federated recommendations~\cite{Yao2021DeviceCloudCL,Qu2024CDCGNNFed} and on-device learning~\cite{Gong2020EdgeRecRS,Han_Ma_Mei_Liu_2021} provide decentralized privacy solutions, they typically incur extra communication overheads and require compatible client-side infrastructure. Our approach, in contrast, targets server-side robustness against missing features, making it orthogonal to distributed training paradigms and directly applicable to standard low-latency serving environments.


\section{Methodology}\label{sec:proposed_framework}
\begin{figure}[t]
    \centering
    \includegraphics[width=\linewidth]{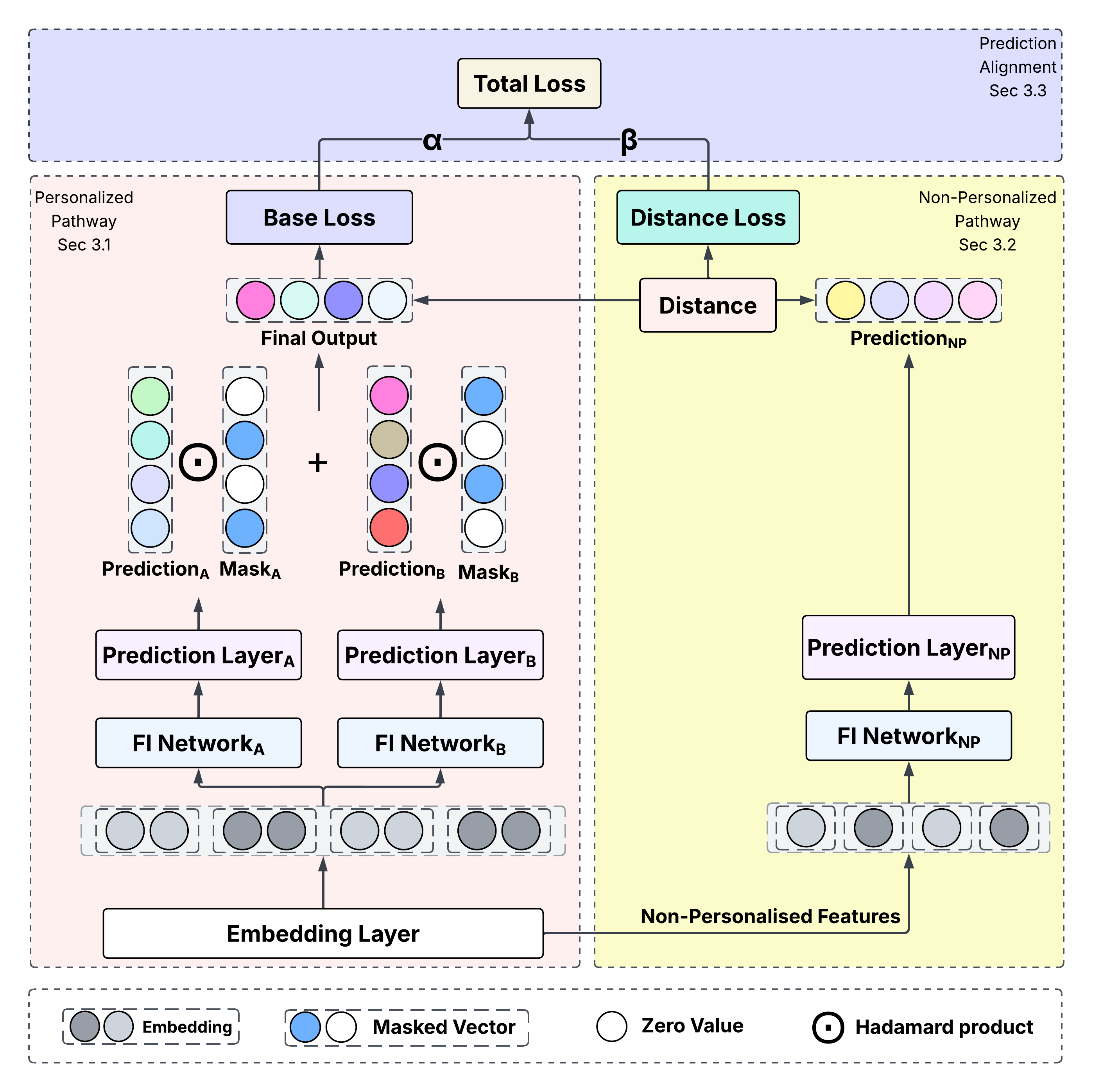}
    \Description{A diagram illustrating the architecture of the proposed RAMP. It shows a personalized pathway with two parallel towers (Tower A and Tower B) sharing an embedding layer. Tower A handles personalized samples, and Tower B handles non-personalized samples using output masking. A separate Non-Personalized pathway trained with only non-personalized features. The two pathways are aligned by a distillation-inspired architecture.}
    \caption{\small The proposed RAMP. The personalized and the non-personalized pathways are trained jointly. A lightweight loss is adopted to enable the distillation-inspired architecture that allows knowledge to be aligned between the two components.}
    \label{fig:FCN-MT_structure}
\end{figure}

The overview of RAMP is shown in Figure~\ref{fig:FCN-MT_structure}. The model consists of two components: (i) a personalized pathway built upon a dual-tower component (pink block), comprising a personalized tower (Tower-A) and a non-personalized tower (Tower-B); (ii) a parallel non-personalized pathway (yellow block), which is trained solely on non-personalized features; and (iii) a distillation-inspired prediction-alignment architecture between (i) and (ii) (purple block). We detail each component in the following subsections.

\subsection{Personalized Pathway}\label{sec:proposed_framework:multitower}

To address the mismatch in feature availability between personalized and non-personalized settings, we introduce a personalized pathway built upon a dual-tower component that separates their learning processes, thereby reducing cross-regime interference. Specifically, both towers use the same input and features for training. However, the prediction of the towers is selectively masked based on personalized and non-personalized traffic. The pathway is model-agnostic and follows the standard design of deep CTR/CVR models, consisting of an embedding layer, a feature interaction (FI) layer, and a prediction layer~\cite{DBLP:conf/wsdm/WangWLGL0G23}.

\subsubsection{Embedding Layer}\label{sec:proposed_framework:multitower:emb_layer}

The embedding layer transforms both dense and sparse (one-hot encoded) features into low-dimensional dense representations. To mitigate data sparsity and promote cross-feature learning across personalized and non-personalized regimes, we adopt a shared embedding layer for all samples. This design is model-agnostic and can be readily integrated with various CTR/CVR backbone models.

\subsubsection{FI Layer}\label{sec:proposed_framework:multitower:fi_layer}

The FI layer captures relationships among features using various interaction modules, such as Wide\&Deep-style models~\cite{Chen_Wang_Liu_Tang_Guo_Zheng_Yao_Zhang_He_2021,gdcn,Wang_Fu_Fu_Wang_2017,Li_Zhang_Zhang_Li_Sang_Zhu_2024} and MLP-based architectures~\cite{Mao_Zhu_Su_Cai_Li_Dong_2023, Zhu_Jia_Cai_Dai_Li_Dong_Tang_Zhang_2023}. 

Following prior work such as~\cite{Lian_Zhou_Zhang_Chen_Xie_Sun_2018}, FI can be broadly categorized as implicit and explicit. Implicit interactions, typically modeled by MLPs, learn high-order feature relationships in a vector-wise manner, where interactions are entangled and not explicitly parameterized (e.g., complex dependencies among user history, context, and item features). In contrast, explicit interactions directly model bounded-order feature crosses in a bit-wise manner, where interactions are explicitly constructed and controlled (e.g., second-order user–item or user–context feature combinations), leading to improved interpretability~\cite{Li_Zhang_Zhang_Li_Sang_Zhu_2024}. By combining both types of interactions, the FI network can capture complementary feature patterns and enhance prediction performance.


The FI network in the personalized pathway consists of two parallel towers, $FI_A$ and $FI_B$, which serve as the backbone model. We adopt independent towers with identical architectures, as this configuration yields the best empirical results. We leave further exploration of architecture and parameter sharing for future work. Both towers use the same embeddings for all features. For data without personalized features, we apply mode-based padding and freeze the weights of the padded embeddings, which preserves a unified input space across samples with different feature availability.

\subsubsection{Prediction Layers}\label{sec:proposed_framework:multitower:pred_layer}
The prediction layers estimate the probability of interaction $\hat{y}$ based on the output of the feature layer. In the proposed design, the prediction layers, $PL_A$ and $PL_B$ (respectively), of the dual-tower component specialize in different types of traffic: personalized and non-personalized (respectively).

More formally, let $ \mathbf{m} = (m_1,\ldots,m_b)$ denote a mask over a batch of $b$ samples, indicating whether $i^{th}$ sample is personalized, i.e., $m_i = 1$, or not $m_i=0$. The prediction of tower A is as follows,
\begin{equation} \label{eqn:personalized_mask}
\begin{split}
\hat{\mathbf{y}}_{A} = \hat{\mathbf{y}}_{PL_{A}} \odot \mathbf{m},
\end{split} 
\end{equation}
where $\hat{\mathbf{y}}_{PL_A} = (\hat{y}_{PL_A}^1, \ldots,\hat{y}_{PL_A}^b )$ represents the prediction made by tower A before masking, and $\odot$ denotes element-wise product.
Conversely, the prediction of tower B handling non-personalized traffic is given by:
\begin{equation} \label{eqn:non_personalized_mask}
\begin{split}
\hat{\mathbf{y}}_{B} = \hat{\mathbf{y}}_{PL_{B}} \odot (\mathbf{1}-\mathbf{m}),
\end{split} 
\end{equation}
where $\mathbf{1}$ denotes a vector whose elements are all equal to $1$.  

The final prediction of the personalized pathway combines the already gated outputs from the personalized ($\hat{\mathbf{y}}_{A}$) and non-personalized ($\hat{\mathbf{y}}_{B}$) towers via a regime-dependent mask-based aggregation, as defined in Equation~\ref{eqn:foundation_prediction}, rather than a generic summation.

\begin{equation}\label{eqn:foundation_prediction}
\hat{\mathbf{y}} = \hat{\mathbf{y}}_{A} + \hat{\mathbf{y}}_{B}.
\end{equation}

Since $\hat{\mathbf{y}}_{A}$ and $\hat{\mathbf{y}}_{B}$ are produced after mask-based gating, their supports are disjoint across traffic regimes, making the above summation equivalent to a selection over personalized and non-personalized inputs without requiring additional normalization to lie within $[0,1]$. At inference time, Tower-A is applied to personalized traffic while Tower-B handles non-personalized traffic within the dual-tower component.

Finally, a loss function is adopted to minimize the difference between the predicted labels $\hat{\mathbf{y}}$ and the true labels $\mathbf{y}$, namely the base loss:
\begin{equation}\label{eqn:base_loss}
\mathcal{L}_{base} = \frac{1}{b} \sum_{i=1}^{b} \ell(\hat{{\mathbf{y}}}_i, {\mathbf{y}}_i),
\end{equation}
where, $\ell(\cdot)$ represents the loss function, and $\mathbf{y}_i$ denotes the ground-truth label. Without loss of generality, in this paper, we adopt the commonly used binary cross-entropy (BCE) loss.

\subsection{Non-Personalized Pathway}\label{sec:proposed_framework:non_per_predict}
To fully explore the scenario where only non-personalized features are available, we introduce a \emph{non-personalized} pathway that takes only non-personalized features as input and produces predictions $\hat{\mathbf{y}}_{np}$ via a dedicated FI network $FI_{NP}$ and the corresponding prediction layer $PL_{NP}$. Non-personalized features can be collected without user consent, as they are not considered sensitive and do not pose a risk of leaking personal data.

As shown in Figure~\ref{fig:FCN-MT_structure}, 
$FI_{NP}$ provides a configurable feature interaction channel to form an additional view over non-personalized inputs.
We train this pipeline \emph{on all samples but using only non-personalized features}. This pathway is trained jointly with the foundational dual-tower component, as detailed in Algorithm \ref{alg:dtcn}. 

\subsection{Distillation-Inspired Prediction-Alignment}\label{sec:proposed_framework:kd_architecture}
With both pathways in place, our goal is to align their behavior using a knowledge-distillation-inspired objective. Let $\hat{\tilde{\mathbf{y}}}_{np}$ and $\hat{\tilde{\mathbf{y}}}$ denote the logits (before the Sigmoid activation) corresponding to the predictions $\hat{\mathbf{y}}_{np}$ and $\hat{\mathbf{y}}$ respectively. We then quantify the discrepancy between the two pathways as follows:

\begin{equation}
    \mathcal{D}_{dis} = \frac{1}{b_{np}}\sum_{i = 1}^{b_{np}} \|{\hat{\tilde{\mathbf{y}}}_i - \hat{\tilde{\mathbf{y}}}_{{np}_{i}}}\|_1,
\end{equation}
where $b_{np}$ is the number of non-personalized samples in a batch. Since CTR/CVR predictions are scalar probabilities in $[0,1]$, the per-sample discrepancy is naturally bounded, making $\mathcal{D}_{dis}$ a stable alignment signal.

\paragraph{Choice of design (loss on the discrepancy).}
A natural option would be to directly minimize the L1 discrepancy between the two pathways. Instead, we apply a cross-entropy-style objective to $\mathcal{D}_{dis}$, name it as the distance loss $\mathcal{L}_{dis}$, which can be interpreted through an \emph{energy-based} view of alignment: distances serve as an ``energy'' measuring mismatch between two views, and logistic / cross-entropy losses provide a smooth mechanism to shape this energy during training~\cite{lecun2006energy_tutorial,hadsell2006invariant_mapping,Hofstaetter2020CrossArchKD}. Empirically, this formulation yielded stronger and more stable improvements on non-personalized traffic, which we attribute to its well-behaved gradients and implicit re-weighting across samples in large-scale noisy recommendation data. This objective is also closely related in spirit to knowledge distillation, especially as the behaviors of the two components are aligning~\cite {DBLP:journals/corr/HintonVD15,Hofstaetter2020CrossArchKD}.

We also explored alternative objectives on the feature level that (i) decouple the semantic spaces of personalized and non-personalized features (field uniformity), or (ii) explicitly align non-personalized features to personalized ones (feature alignment). Empirically, both were less effective, suggesting that knowledge-distillation-inspired actions on prediction logits is a more reliable mechanism. One possible explanation is that both objectives intervene directly in the representation space, either encouraging excessive separation or enforcing overly strict correspondence between the two pathways. In contrast, logit-level alignment regularizes only the final predictive behavior, preserving the flexibility of each pathway to learn task-specific representations while still transferring useful knowledge.

\begin{algorithm}[ht]
\caption{RAMP Training Algorithm}
\label{alg:dtcn}
\begin{algorithmic}[1]
\Require Training dataset $\mathcal{D} = \{(\bx_i, y_i, m_i)\}_{i=1}^{N}$, learning rate $\eta$, batch size $b$, number of non-personalized samples in a batch $b_{np}$, embedding layer $\mathcal{E}$, feature interaction networks $FI_A$, $FI_B$, $FI_{NP}$, prediction layers $PL_A$, $PL_B$, $PL_{NP}$
\Ensure Optimized model parameters $\Theta$
\For{each epoch}
    \For{each mini-batch $\mathcal{B} = \{(\bx_i, y_i, m_i)\}_{i=1}^{b}$ in $\mathcal{D}$}
        \State $\mathbf{E} \gets \mathcal{E}(\bx_i)$ for all $i \in \mathcal{B}$ \Comment{Embedding Layer}
        
        \State $\mathbf{H}_A \gets FI_A(\mathbf{E})$ \Comment{Tower A: input processing}
        \State $\mathbf{H}_B \gets FI_B(\mathbf{E})$ \Comment{Tower B: input processing}
        
        \State $\hat{\by}_{PL_A} \gets PL_A(\mathbf{H}_A)$ \Comment{Tower A Prediction}
        \State $\hat{\by}_{PL_B} \gets PL_B(\mathbf{H}_B)$ \Comment{Tower B Prediction}
        \State $\bmm \gets (m_1, m_2, \ldots, m_b)^T$ \Comment{Personalized mask} 
        \State $\hat{\by}_A \gets \hat{\by}_{PL_A} \odot \bmm$ \Comment{Tower A masked prediction}
        \State $\hat{\by}_B \gets \hat{\by}_{PL_B} \odot (\mathbf{1} - \bmm)$ \Comment{Tower B masked prediction}
        
        \State $\hat{\by} \gets \hat{\by}_A + \hat{\by}_B$ \Comment{Combined prediction}
        \State $\bx^{np}_i \gets \bx_i$ if $m_i = 0$ \Comment{Non-PER samples}
        \State $\mathbf{E}^{np} \gets \mathcal{E}(\bx^{np}_i)$ \Comment{Reuse embedding layer}
        \State $\mathbf{H}_{NP} \gets FI_{NP}(\mathbf{E}^{np})$ \Comment{Separate FI for knowledge distillation}
        \State $\hat{\by}_{np} \gets PL_{NP}(\mathbf{H}_{NP})$
        \State $\mathcal{D}_{\text{dis}} \gets \frac{1}{b_{np}}\sum_{i=1}^{b_{np}} \|\hat{\tilde{y}}_i - \hat{\tilde{y}}_{np,i}\|_1$ \Comment{L1 distance}
        \State $\mathcal{L}_{\text{base}} \gets \frac{1}{b} \sum_{i=1}^{b} \text{BCE}(\hat{y}_i, y_i)$ 
        \State $\mathcal{L}_{\text{overall}} \gets \alpha \cdot \mathcal{L}_{\text{base}} + \beta \cdot \mathcal{L}_{\text{dis}}$
    
        \State Update all parameters: $\Theta \gets \Theta - \eta \cdot \nabla_{\Theta} \mathcal{L}_{\text{overall}}$
    \EndFor
\EndFor
\State \Return Trained model parameters $\Theta$
\end{algorithmic}
\end{algorithm}

\subsection{The Combined Loss}
We combine the distillation-inspired prediction-alignment architecture to the foundational CTR/CVR prediction task through joint optimization of an overall loss, defined as:

\begin{equation}
    \mathcal{L}_{overall} = \alpha \cdot \mathcal{L}_{base} + \beta \cdot \mathcal{L}_{dis},
\end{equation}
where $\mathcal{L}_{overall}$ denotes the overall loss. $\alpha$ and $\beta$ are assigned to $\mathcal{L}_{base}$ and $\mathcal{L}_{dis}$ as hyperparameters to control the relative importance of loss components. $\beta$ is 0 if the prediction-alignment architecture is not activated. For the simplicity of hyper-parameter search, we fix the value of $\alpha$ to 1, and control the relative importance between two loss components by altering the value of $\beta$. 
We use logit-level alignment instead of feature-level alignment because the two pathways rely on different feature sets. Feature-level alignment may force incompatible representations to become similar, weakening pathway specialization. Logit-level alignment transfers predictive behavior while preserving feature-specific representations. Algorithm~\ref{alg:dtcn} depicts the full training procedure of the proposed RAMP model.

\begin{table*}[htb]
    \centering
    \small
    \caption{Statistics of datasets. \#Features is the total count of unique feature values. \#PER.Fields and \#Non-PER.Fields represent the number of personalized (PER) and non-personalized (Non-PER) feature fields with corresponding examples followed. }
    \label{tab:dataset_statistics}
    \begin{tabular}{l|ccccccc}
    \hline
    \textbf{Dataset} & \textbf{Task} & \textbf{\#Samples} & \textbf{\#Features} & \textbf{\#PER.Fields} & \textbf{PER.Example} & \textbf{\#Non-PER.Fields} & \textbf{Non-PER.Example}\\
    \hline
    Avazu           & CTR & 40,428,967  & 3,750,999 & 5  & device\_ip     & 19 & app\_id       \\
    TaobaoAd        & CTR & 26,557,961  & 1,295,076 & 12 & userid         & 8  & adgroup\_id   \\
    CriteoPrivateAd & CVR & 103,862,032 & 4,015,683 & 32 & user\_id       & 18 & campaign\_id  \\
    IndustryAd      & CVR & 27,424,619  & 1,229,008 & 30 & app\_click\_id & 30 & site\_id      \\
    \hline
    \end{tabular}
\end{table*}
\section{Experiments} 
In this section, we present experimental results on real-world datasets for both CTR and CVR prediction tasks, and answer the following research questions.
\begin{itemize}[leftmargin=*]
  \item \textbf{RQ1:} How does RAMP compare to established deep CTR/CVR baseline models on prediction accuracy, especially for data without personalized features available?
  \item \textbf{RQ2:} Do the key components of the proposed RAMP effectively contribute to the improvement in performance?
  \item \textbf{RQ3:} Can our proposed RAMP framework learn effective representations of personalized and non-personalized features?
\end{itemize}

\label{sec:experiments}
\subsection{Experimental Setup}
\label{sec:experiments_setup}

\subsubsection{Datasets}
\label{sec:experiments:setup:datasets}
We conduct extensive experiments on four real-world datasets, as summarized in Table~\ref{tab:dataset_statistics}, including: (i) two public CTR datasets (Avazu \citep{avazu2014} and TaobaoAd \citep{DSIN}); and (ii) two CVR datasets (CriteoPrivateAd\footnote{ From \url{https://huggingface.co/datasets/criteo/CriteoPrivateAd}.} \citep{sebbar2025criteoprivateads} and a private industrial dataset IndustryAd). These datasets span diverse scales and feature compositions, with varying numbers of personalized and non-personalized feature fields as detailed in Table~\ref{tab:dataset_statistics}, enabling comprehensive evaluation of our proposed framework across different prediction tasks. 

\subsubsection{Data Preprocessing}
We adopt a well-established methodology~\cite{DBLP:conf/cikm/ZhuLYZH21} widely used for benchmarking CTR/CVR prediction models to preprocess all four datasets. 
The preprocessing pipeline consists of standard procedures: categorical features are mapped to unique integer IDs and subsequently embedded into dense vectors, while numerical features undergo standard normalization to mitigate the impact of outliers. To manage feature sparsity and reduce dimensionality, infrequent categorical features appearing below a certain threshold are mapped to a special ``Out-of-Vocabulary'' (OOV) token. Specifically, this frequency threshold is set to 10 for the TaobaoAd, CriteoPrivateAd, and IndustryAd datasets, and 2 for the Avazu dataset. 
Each dataset is partitioned into training, validation, and test sets to ensure robust evaluation. For Avazu and TaobaoAd, we adhere to the predefined splits from the original datasets. For CriteoPrivateAd and our private IndustryAd dataset, we perform a chronological split based on timestamps to simulate the real-world deployment scenario and prevent data leakage.

To systematically evaluate model prediction performance under personalized-feature constraints, we construct both personalized and non-personalized versions of the three public datasets (Avazu, TaobaoAd, and CriteoPrivateAd) based on the feature categorization presented in Table~\ref{tab:dataset_statistics}. For each dataset, we create a \textit{pure non-personalized dataset} by masking all personalized feature fields, while retaining only non-personalized features. This masking procedure is consistently applied to training, validation, and test sets. The original datasets, containing both personalized and non-personalized features, serve as the \textit{personalized datasets}. For all experiments reported in this section, we utilize the \textit{mixed dataset} that combines samples from both the personalized and non-personalized datasets. In contrast, the private IndustryAd dataset naturally contains a realistic mixture in production environments, with 20\% of samples having missing personalized features (non-personalized traffic) and 80\% containing complete personalized features, which can be used directly without additional masking. 


We acknowledge that feature masking simulates feature \emph{absence} rather than fully capturing the potential distributional differences between consenting and non-consenting users. In practice, users who decline consent may exhibit behavioral patterns that differ from those of users with available personalized features. Following prior work, we adopt feature masking as a controlled protocol to isolate the impact of feature availability on model performance. To partially bridge the gap to real-world deployments, we additionally evaluate on the IndustryAd dataset, which contains a naturally occurring 80:20 traffic split without synthetic masking. The consistent improvements observed on both the public benchmark datasets and IndustryAd suggest that RAMP remains effective under both synthetic and naturally occurring feature-availability constraints (see Table~\ref{tab:main_results}), although exploring more pronounced distribution shifts remains an important direction for future work.

\subsubsection{Baselines}
To demonstrate the effectiveness of RAMP, we compare its performance against a comprehensive set of state-of-the-art deep learning models for CTR/CVR prediction. These baselines represent various architectural designs and feature interaction mechanisms. We include MLP-based models such as DNN~\cite{covington2016deep} and FINAL~\cite{Zhu_Jia_Cai_Dai_Li_Dong_Tang_Zhang_2023}, which primarily rely on multi-layer perceptrons. We also compare our approach with FM-based models like DeepFM~\cite{Guo_Tang_Ye_Li_He_2017} and xDeepFM~\cite{Lian_Zhou_Zhang_Chen_Xie_Sun_2018}, which combine the strengths of factorization machines with deep neural networks for high-order feature interactions. Furthermore, we select several high-performance and even higher-order interaction-focused models, including PNN~\cite{DBLP:conf/icdm/QuCRZYWW16}, Wide \& Deep~\cite{wide_deep_2016}, DCN~\cite{Wang_Fu_Fu_Wang_2017}, AutoInt~\cite{song2019autoint}, AFN~\cite{cheng2020adaptive}, FiBiNet~\cite{fibinet}, EDCN~\cite{Chen_Wang_Liu_Tang_Guo_Zheng_Yao_Zhang_He_2021}, GDCN~\cite{gdcn}, EulerNet~\cite{tian2023eulernet}, DCNv2~\cite{dcnv2}, MaskNet~\cite{Wang_She_Zhang_2021}, and FCN~\cite{Li_Zhang_Zhang_Li_Sang_Zhu_2024}, which introduce various mechanisms (including and not limited to gate control and adaptive feature importance learning) to capture complex feature interactions. Although these models are not inherently privacy-preserving, we adapt them for non-personalized evaluation by masking personalized inputs, establishing them as robust non-private baselines. To evaluate the performance of RAMP, we compared it with existing knowledge distillation baselines, which will be discussed later in Section~\ref{sec:kd_comparison}.
\begin{table*}[h]
    \centering
    \small
    \caption{\small Non-PER performance comparison (LogLoss↓, AUC↑). PP/RAMP use the backbone that yields the best PP/RAMP result on each dataset (PNN for TaobaoAd/CriteoPrivateAd; FCN for Avazu/IndustryAd). $\Delta_{abs}$ and $\Delta_{rel}$ report RAMP’s improvement over the best baseline. $^{*}$ indicates a statistically significant difference between RAMP and the best baseline under a two-tailed t-test at significance level $\alpha=0.001$. RAMP significantly outperforms all baselines.}
    \label{tab:main_results}
    \begin{tabular}{l|cc|cc|cc|cc}
    \hline
    \multirow{2}{*}{Models} & \multicolumn{2}{c|}{Avazu} & \multicolumn{2}{c|}{TaobaoAd} & \multicolumn{2}{c|}{CriteoPrivateAd} & \multicolumn{2}{c}{IndustryAd} \\
    \hhline{~|--|--|--|--}
    & LogLoss↓ & AUC(\%)↑ & LogLoss↓ & AUC(\%)↑ & LogLoss↓ & AUC(\%)↑ & LogLoss↓ & AUC(\%)↑ \\
    \hline
    DNN~\cite{covington2016deep} & 0.3934 & 75.42 & 0.1978 & 58.55 & 0.4209 & 77.70 & 0.0884 & 80.17 \\
    PNN~\cite{DBLP:conf/icdm/QuCRZYWW16} & 0.3929 & 75.50 & \underline{0.1970} & \underline{59.17} & 0.4136 & 78.17 & 0.0894 & 80.15\\
    Wide \& Deep~\cite{wide_deep_2016} & 0.3933 & 75.43 & 0.1989 & 58.50 & 0.4137 & 78.19 & 0.0884 & 79.76  \\
    DeepFM~\cite{Guo_Tang_Ye_Li_He_2017} & 0.3932 & 75.45 & 0.1994 & 58.37 & 0.4140 & 78.37 & 0.0916 & 80.48 \\
    DCN~\cite{Wang_Fu_Fu_Wang_2017} & 0.3933 & 75.45 & 0.1982 & 58.48 & 0.4133 & 78.40 & 0.0982 & 79.61\\
    xDeepFM~\cite{Lian_Zhou_Zhang_Chen_Xie_Sun_2018} & 0.3934 & 75.44 & 0.1984 & 58.11 & 0.4130 & 78.12 & 0.0875 & 80.83  \\
    AutoInt~\cite{song2019autoint} & 0.3933 & 75.43 & 0.1976 & 58.42 & 0.4146 & 78.19 & 0.0888 & 80.03 \\
    AFN~\cite{cheng2020adaptive} & 0.3926 & 75.56 & 0.1977 & 58.47 & 0.4135 & 78.37 & 0.1188 &72.69 \\
    FiBiNet~\cite{fibinet} & 0.3932 & 75.45 & 0.1983 & 58.31 & 0.4155 & 78.19 & 0.0891 & 80.50 \\
    EDCN~\cite{Chen_Wang_Liu_Tang_Guo_Zheng_Yao_Zhang_He_2021} & 0.3916 & 75.72 & 0.1976 & 58.50 & 0.4134 & 78.32 & 0.0872 & 80.93  \\
    GDCN~\cite{gdcn} & 0.3936 & 75.42 & 0.2005 & 58.78 & 0.4123 & 78.39 & \underline{0.0867} & \underline{81.38} \\
    EulerNet~\cite{tian2023eulernet} & 0.3923 & 75.60 & 0.1979 & 58.33 & 0.4127 & 78.12  & 0.0898 & 80.36  \\
    DCNv2~\cite{dcnv2} & 0.3928 & 75.52 & 0.1978 & 58.60 & 0.4104 & 78.60  & 0.0873 & 81.16  \\
    MaskNet~\cite{Wang_She_Zhang_2021} & 0.3928 & 75.55 & 0.2005 & 58.49 & 0.4131 & 78.46  & 0.0887 & 80.52 \\
    FINAL~\cite{Zhu_Jia_Cai_Dai_Li_Dong_Tang_Zhang_2023} & \underline{0.3916} & \underline{75.74}  & 0.1974 & 58.94 & \underline{0.4097} & 78.55 & 0.0901 & 80.27 \\
    FCN~\cite{Li_Zhang_Zhang_Li_Sang_Zhu_2024} & 0.3926 & 75.56 & 0.1980 & 58.79 & 0.4101 & \underline{78.65} & 0.0870 & 81.12  \\
    \hline
    PP      & 0.3911 & 75.82 & 0.1968 & 59.33 & 0.4098 & 79.03  & 0.0866   & 81.39  \\
    RAMP         & \textbf{0.3910*} & \textbf{75.84*} & \textbf{0.1968*} & \textbf{59.39*} & \textbf{0.4035*} & \textbf{79.52*} & \textbf{0.0865*}   & \textbf{81.49*}    \\
    \hline
    $\Delta_{abs}$ & \textit{-0.0006} & \textit{+0.10} & \textit{-0.0002} & \textit{+0.22} & \textit{-0.0066} & \textit{+0.87} & \textit{-0.0002} & \textit{+0.11} \\
    $\Delta_{rel}$ (\%) & \textit{-0.15} & \textit{+0.13} & \textit{-0.10} & \textit{+0.37} & \textit{-1.61} & \textit{+1.11} & \textit{-0.23} & \textit{+0.14} \\
    \hline
    \end{tabular}
\end{table*}

\subsubsection{Implementation Details} \label{subsec:implementation_details}
We implement all models using PyTorch~\cite{paszke2019pytorch} based on the FuxiCTR framework and benchmarking methods~\cite{DBLP:conf/sigir/ZhuDSMLCXZ22,DBLP:conf/cikm/ZhuLYZH21}. We employ the Adam~\cite{DBLP:journals/corr/KingmaB14} to optimize all models, with a default learning rate set to 0.001. To prevent overfitting, we employ early stopping with a patience value of 2. The embedding dimension is tuned per dataset and per baseline model based on validation performance: 32 is generally best for the TaobaoAd dataset and 16 for the remaining datasets. The batch size is set to 10,000 for all datasets after tuning, as it achieves the optimal performance. Hyperparameters for all baseline models are configured and fine-tuned following the optimal values reported in their original publications and established benchmarks. The proposed framework is evaluated in three configurations based on the backbone model: PP (the foundational personalized pathway) and RAMP (the full framework incorporating knowledge distillation). Regarding the choice of backbone model, we select a backbone based on validation performance for each dataset, and evaluate both Personalized Pathway and RAMP using the same backbone; robustness across different backbones is reported separately: PNN for TaobaoAd and CriteoPrivateAd, and FCN for Avazu and IndustryAd. A comparison across different backbones (PNN, FCN, FINAL) is provided in Section~\ref{sec:impact_of_backbone_models}. Notably, during validation and testing, only the foundational dual-tower component is used for prediction. The non-personalized pathway (Section~\ref{sec:proposed_framework:non_per_predict}) and the distillation-inspired prediction-alignment architecture (Section~\ref{sec:proposed_framework:kd_architecture}) are only activated during training. The non-personalized pathway is thus not required during inference, and its benefits are incorporated into the personalized pathway as the behavior of these two pathways is aligned. All experiments are conducted on a single NVIDIA Tesla H100 GPU with five fixed random seeds for reproducibility. The training cost of the method is introduced in Appendix~\ref{sec:training_efficiency}.

\subsubsection{Evaluation Metrics}
Following the practice of the community, we employ two metrics: LogLoss and Area Under the ROC Curve (AUC)~\cite{Yang_Pan_2022}. Following prior work~\cite{dcnv2, song2019autoint}, even a 0.1\% improvement in AUC can have a substantial practical impact in computational advertising at scale, provided that the gain is statistically supported. To ensure reliable evaluation, all results are averaged over 5 runs with different random seeds. Since our training data contains both personalized and non-personalized samples, as described in Section~\ref{sec:experiments:setup:datasets}, we evaluate model performance separately on these two subsets to assess their effectiveness under different privacy scenarios. Specifically, we report performance from two perspectives: 1) the performance on the non-personalized subset using non-personalized features only, evaluated with the entire architecture, denoted as Non-PER AUC and Non-PER LogLoss; and 2) the performance on the personalized subset with all features, evaluated with the entire architecture, denoted as PER AUC and PER LogLoss. \textbf{Our primary focus is on the non-personalized performance}, as it represents the most challenging and practically relevant scenario for privacy-preserving advertising systems.

\subsubsection{Privacy Compliance}
RAMP is designed to support deployment under consent-driven feature restrictions, in line with data protection regulations. During online inference (only the dual-tower component is used during inference), the system adapts to feature availability: when personalized features are available (i.e., consented traffic), both pathways are used; when such features are unavailable, only the non-personalized pathway is activated, relying exclusively on non-user-linked attributes (e.g., context, ad properties). Prediction alignment is applied only during training on data with available personalized features, enabling alignment without requiring access to personalized data during inference on non-consented traffic.

\begin{table}[h]
    \centering
    \small
    \caption{Performance on personalized data (PER AUC) for representative baseline models using all features.}
    \label{tab:PER_results}
    \begin{tabular}{l|cccc}
    \hline
    Models & Avazu & TaobaoAd & CriteoPrivateAd & IndustryAd \\
    \hline
    PNN          & 79.69 & 65.14 & 81.51 & 83.60 \\
    DCNv2        & 79.52 & 64.96 & 81.95 & 84.29 \\
    FINAL        & 79.53 & 64.81 & 81.89 & 83.66 \\
    FiBiNet      & 79.26 & 64.86 & 81.69 & 83.60 \\
    MaskNet      & 79.54 & 64.78 & 81.92 & 83.77 \\
    FCN          & 79.86 & 65.04 & 82.09 & 84.38 \\
    \hline
    RAMP         & 79.83 & 65.18 & 82.10 & 84.26 \\
    \hline
    \end{tabular}
\end{table}
\subsection{Main Results (RQ1)}
\label{sec:main_results}

Table~\ref{tab:main_results} shows the comprehensive non-personalized performance comparison, which is our main focus of evaluation. Among the baseline models, we observe considerable performance variations across datasets. FINAL, PNN and FCN demonstrate relatively competitive performance on smaller datasets (Avazu and TaobaoAd), while GDCN and FCN achieve the best baseline results on larger datasets (IndustryAd and CriteoPrivateAd). Notably, simpler architectures like PNN occasionally outperform more complex models, suggesting that model complexity does not always translate to better non-personalized performance. This observation highlights the inherent challenge of non-personalized prediction, where conventional feature interaction mechanisms struggle to compensate for missing personalized signals, motivating our dual-tower component design with explicit prediction alignment.

PP consistently improves upon the best baseline models across all datasets, with AUC improvements of up to +0.38\% (on CriteoPrivateAd), validating the effectiveness of the dual-tower component in learning from mixed personalized and non-personalized training data. Building upon this foundation, RAMP achieves further improvements, obtaining the best results among the evaluated methods with AUC gains of +0.10\% to +0.87\% over previous best models. Particularly noteworthy, we achieve the most significant performance gain on the CriteoPrivateAd dataset (+0.87\% AUC), indicating that the proposed method remains effective even in large-scale scenarios. All improvements are statistically significant ($p < 1e-3$) and exceed the 0.1\% threshold considered meaningful in computational advertising~\cite{song2019autoint}. Moreover, RAMP was A/B tested in industry production for CVR prediction, the total advertiser value (TAV) improved over 3\% (See Appendix~\ref{sec:ab_test}). 

We also evaluate the performance of representative baseline models on personalized data (PER AUC), where all models have access to complete personalized features. As shown in Table~\ref{tab:PER_results}, baseline models achieve reasonable performance on personalized data, with FCN demonstrating competitive results across most datasets. 

Overall, these results demonstrate that RAMP improves prediction under personalized-feature constraints by enabling the non-personalized pathway to effectively learn from the dual-tower foundation, yielding strong gains on non-personalized AUC while remaining competitive on personalized AUC.

\vspace{-0.4cm}
\subsection{Impact of Different Components (RQ2)}
\label{sec:ablation_study}

\subsubsection{Effect of Our Architecture Across Different Backbones}\label{sec:impact_of_backbone_models}
Given the model-agnostic nature of RAMP, we seek to investigate the effect of our architecture across different backbones. We use TaobaoAd and CriteoPrivateAd for this experiment. One is a CTR dataset with a relatively balanced personalized-to-non-personalized feature ratio, while the other is a large-scale CVR dataset. The results are reported in Table~\ref{tab:model_variants_results}. Looking at the baseline models (backbone-only), PNN excels on TaobaoAd and FCN on CriteoPrivateAd. The dual-tower component consistently improves all backbones, with improvement magnitudes varying by model capacity---PP-PNN gains +0.64\% AUC on CriteoPrivateAd while PP-FCN gains +0.38\%, suggesting the component particularly benefits models with initially limited non-personalized performance. The complete RAMP framework further improves overall performance-- RAMP-PNN gains +1.39\% AUC on CriteoPrivateAd compared to PNN-Base and +0.75\% AUC over PP-PNN.
\begin{table}[tb]
    \centering
    \small
    \caption{\small Non-PER performance comparison of RAMP with different backbone models using only non-personalized features. Base denotes the original baseline model, PP denotes using the personalized pathway only, and RAMP denotes the full framework.}
    \label{tab:model_variants_results}
    \begin{tabular}{l|cc|cc}
    \hline
    \multirow{2}{*}{Models} & \multicolumn{2}{c|}{TaobaoAd} & \multicolumn{2}{c}{CriteoPrivateAd} \\
    \hhline{~|--|--}
               & LogLoss↓ & AUC(\%)↑ & LogLoss↓ & AUC(\%)↑ \\
    \hline
    PNN-Base   & 0.1970 & 59.17 & 0.4146 & 78.13 \\
    PP-PNN     & 0.1968 & 59.33 & 0.4087 & 78.77 \\
    RAMP-PNN   & 0.1968 & 59.39 & 0.4035 & 79.52 \\
    \hline
    FCN-Base   & 0.1980 & 58.79 & 0.4101 & 78.65 \\
    PP-FCN     & 0.1972 & 58.94 & 0.4098 & 79.03 \\
    RAMP-FCN   & 0.1969 & 59.28 & 0.4073 & 79.21 \\
    \hline
    FINAL-Base & 0.1974 & 58.94 & 0.4097 & 78.55 \\
    PP-FINAL   & 0.1972 & 59.17 & 0.4084 & 78.83 \\
    RAMP-FINAL & 0.1970 & 59.29 & 0.4079 & 78.89  \\
    \hline
    \end{tabular}
\end{table}

\begin{table}[b]
    \centering
    \small
    \caption{\small Comparison across knowledge distillation baselines on Non-PER performance. Best results \textbf{bolded}, second-best \underline{underlined}. }
    \label{tab:kd_comparison}
    \begin{tabular}{l|cc|cc}
    \hline
    \multirow{2}{*}{Method} & \multicolumn{2}{c|}{Avazu} & \multicolumn{2}{c}{TaobaoAd} \\
    \hhline{~|--|--}
    & LogLoss↓ & AUC(\%)↑ & LogLoss↓ & AUC(\%)↑  \\
    \hline
    Backbone                               & 0.3916 & 75.74 & 0.1970 & 59.17 \\
    KD~\cite{DBLP:journals/corr/HintonVD15}& 0.3913 & 75.78 & 0.1969 & 59.27 \\
    PFD~\cite{DBLP:conf/kdd/XuLGGYPSWSO20} & 0.3914 & 75.78 & 0.1969 & 59.29 \\
    HAPFD~\cite{DBLP:conf/kdd/Yuan0HD25}   & \underline{0.3913} & \underline{75.80} & \underline{0.1968} & \underline{59.30} \\
    \hline
    RAMP       & \textbf{0.3910} & \textbf{75.84} & \textbf{0.1968} & \textbf{59.39} \\
    \hline
    \end{tabular}
\end{table}

\subsubsection{Comparison with Knowledge Distillation Baselines}
\label{sec:kd_comparison}
To further validate the effectiveness of RAMP's prediction alignment module, we compare it against representative knowledge distillation methods, all integrated into the same dual-tower component to ensure a fair comparison. All methods share the same backbone (FINAL for Avazu, PNN for TaobaoAd), training data, student input (non-personalized features only), and optimization budget; they differ only in the distillation objective used to train the non-personalized student model. The compared methods are: (1)~\textbf{KD}~\cite{DBLP:journals/corr/HintonVD15}, which applies standard KL-divergence distillation from the dual-tower teacher; (2)~\textbf{PFD}~\cite{DBLP:conf/kdd/XuLGGYPSWSO20}, which employs privileged feature distillation with feature-level alignment; and (3)~\textbf{HAPFD}~\cite{DBLP:conf/kdd/Yuan0HD25}, which extends PFD with heterogeneous-aware focal weighting to rebalance distillation across sample types.

\begin{figure*}[tb]
  \centering
  \begin{subfigure}[t]{0.66\linewidth}
    \centering
    \includegraphics[height=3.1cm]{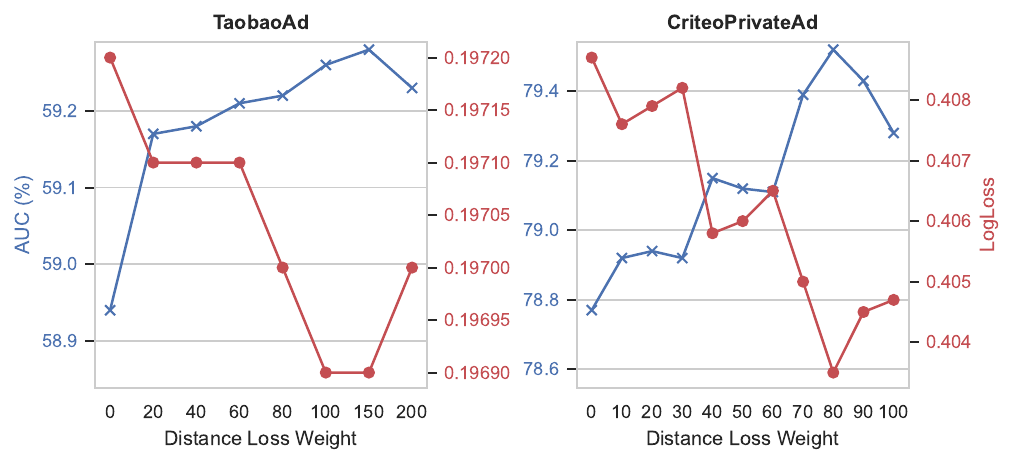}
    \caption{}
    \label{fig:distance_hyper}
  \end{subfigure}
  \hfill
  \begin{subfigure}[t]{0.33\linewidth}
    \centering
    \includegraphics[height=3.1cm]{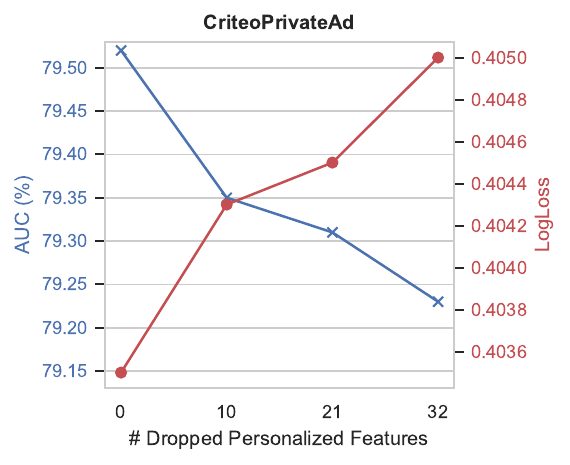}
    \caption{}
    \label{fig:feature_ablation}
  \end{subfigure}
  \caption{\small Ablation results for RAMP. (a) Effect of the distance-loss weight $\beta$ on Non-PER performance: Non-PER AUC (blue crosses) and LogLoss (red dots) for RAMP-FCN on TaobaoAd (left) and RAMP-PNN on CriteoPrivateAd (right) across $\beta$ values. (b) Effect of dropping an increasing number of personalized feature fields on CriteoPrivateAd: Non-PER AUC (blue crosses) and LogLoss (red dots) for RAMP-PNN.}
  \Description{Two sets of ablation study plots. (a) shows the impact of the distance loss weight beta on model performance (AUC and LogLoss) for TaobaoAd and CriteoPrivateAd datasets, exhibiting an inverted-U trend where performance peaks at an optimal weight range. (b) is a line plot showing the robustness of the model as personalized features are progressively dropped, showing only a slight decline in performance even as the number of features is significantly reduced.}
\end{figure*}
As shown in Table~\ref{tab:kd_comparison}, RAMP consistently outperforms all KD baselines on both datasets. On Avazu, RAMP achieves 75.84\% AUC compared to 75.80\% for the next-best method (HAPFD); on TaobaoAd, RAMP reaches 59.39\% AUC versus 59.30\% for HAPFD. We report results on Avazu and TaobaoAd as they use different backbone configurations (FINAL and PNN respectively), providing evidence of generalization across architectures; extending this comparison to additional datasets is left for future work.

\subsubsection{Sensitivity to Distance Loss Weight}
\label{sec:distance_loss_weight_hyper}

To investigate the sensitivity of RAMP to the distance loss weight hyperparameter $\beta$, we conduct experiments using RAMP-FCN on TaobaoAd and RAMP-PNN on CriteoPrivateAd, as illustrated in Figure~\ref{fig:distance_hyper}. These experiments are based on the best-performing PP models identified through a hyperparameter search. We adopt a coarse-to-fine search strategy to identify the optimal $\beta$ values, which first identify the approximate effective range for each dataset, then perform fine-grained weight tuning within these ranges as shown in Figure~\ref{fig:distance_hyper}. The results reveal a consistent inverted-U pattern, attributed to the trade-off between the effectiveness of prediction alignment and training stability. We select these two combinations as they correspond to the best-performing PP configurations on each dataset as discussed in Table~\ref{tab:model_variants_results}.
When $\beta$ is too small (e.g., 0-20), the distilled knowledge provides insufficient guidance for the non-personalized pipeline to effectively learn from the dual-tower foundation, resulting in limited knowledge transfer. As $\beta$ increases to the optimal range (80-150), the knowledge learned by the personalized pathway is successfully distilled and transferred, enabling the model to learn better feature representations for non-personalized data. However, when $\beta$ becomes excessively large (>150), the distance loss dominates the overall training objective, potentially causing the model to overfit on the knowledge distillation task at the expense of the primary prediction objective, leading to performance degradation. The final $\beta$ values, selected at the peak points in Figure~\ref{fig:distance_hyper}, are 150 for TaobaoAd and 80 for CriteoPrivateAd.

\subsection{Robustness to Per. Feature Availability}

To evaluate RAMP's robustness under varying degrees of personalized feature availability, we conduct a feature ablation study on CriteoPrivateAd, which contains the largest number of personalized features (32 fields) among the evaluated datasets. We select the best-performing RAMP-PNN model from Section~\ref{sec:distance_loss_weight_hyper} and progressively remove personalized features during \emph{training} to observe the Non-PER performance.
Rather than removing features randomly, we adopt an importance-based removal strategy. 
Feature importance is determined by training the PNN backbone on personalized data and measuring the AUC degradation when each personalized feature is individually removed; features causing larger drops are considered more important. We use the PNN backbone for this ranking as it is the same architecture underlying RAMP-PNN. Features are then removed in descending order of importance. Crucially, this experiment assesses how the quality of the training-time teacher signal from the personalized tower degrades as fewer personalized features are available; the Non-PER inference pathway itself always uses only non-personalized features.

As illustrated in Figure~\ref{fig:feature_ablation}, RAMP-PNN demonstrates remarkable resilience to personalized feature reduction. Starting from a baseline performance of 79.52\% AUC with all 32 features, removing the top 10 most important features (keep 22 personalized features) results in a modest decline to 79.35\% AUC (0.17\% decrease). Even when 21 features are removed (only 11 remained), the model maintains competitive performance at 79.31\% AUC (0.21\% degradation). 
This robustness can be attributed to the synergistic design of RAMP. First, the dual-tower component enables simultaneous learning from both personalized and non-personalized training samples, allowing the non-personalized tower to develop robust representations independent of personalized feature availability. Second, the two-pathway architecture facilitates knowledge distillation from the remaining personalized features to the non-personalized pipeline, effectively compensating for the reduced personalized signal. 

\begin{figure}[h]
  \centering
  \includegraphics[width=0.6\linewidth]{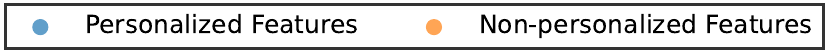}

  \begin{subfigure}[b]{0.45\linewidth}
    \centering
    \includegraphics[width=\linewidth]{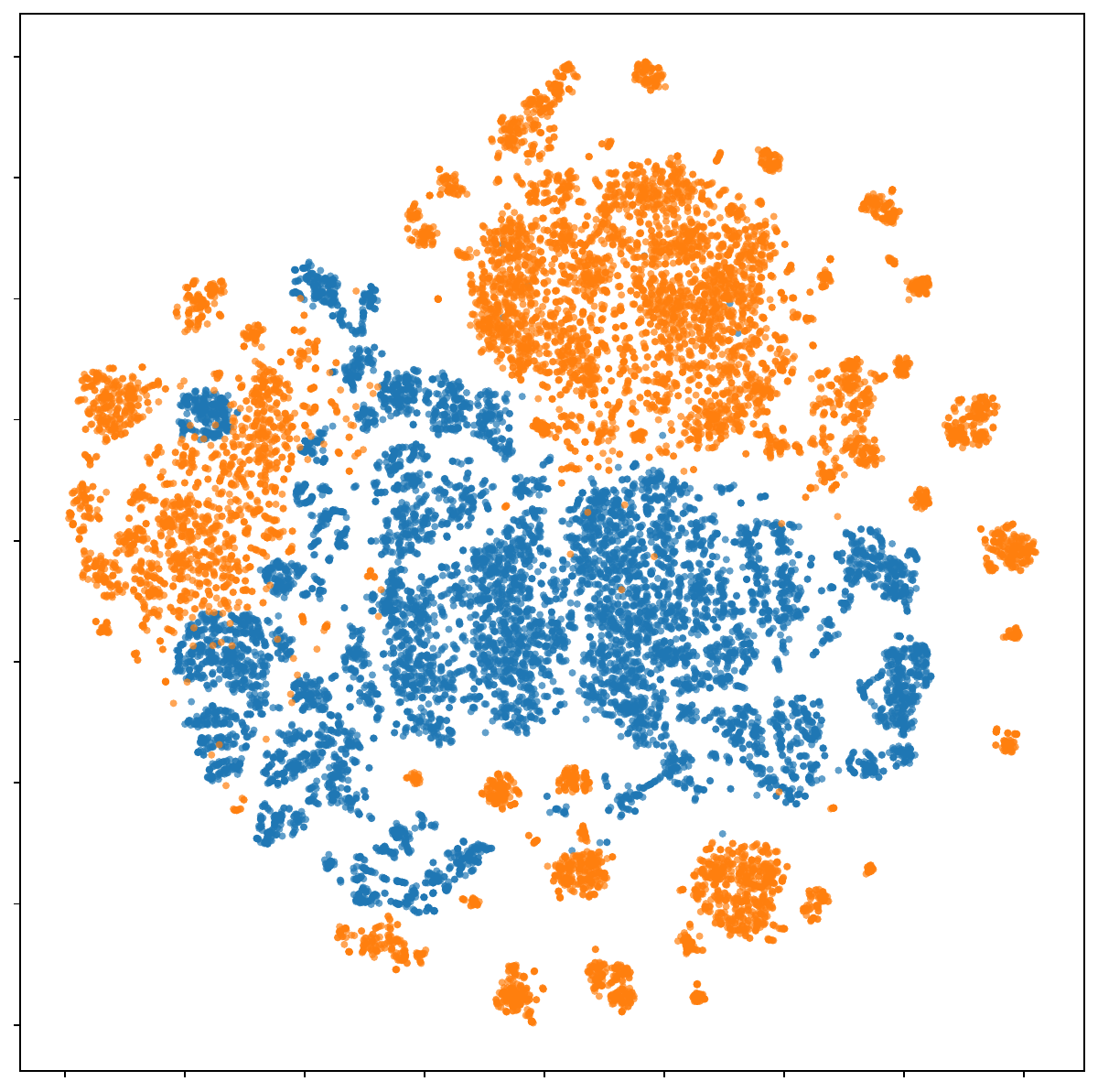}
    \caption{PNN}
    \label{fig:pnn_base}
  \end{subfigure}
  \hfill
  \begin{subfigure}[b]{0.45\linewidth}
    \centering
    \includegraphics[width=\linewidth]{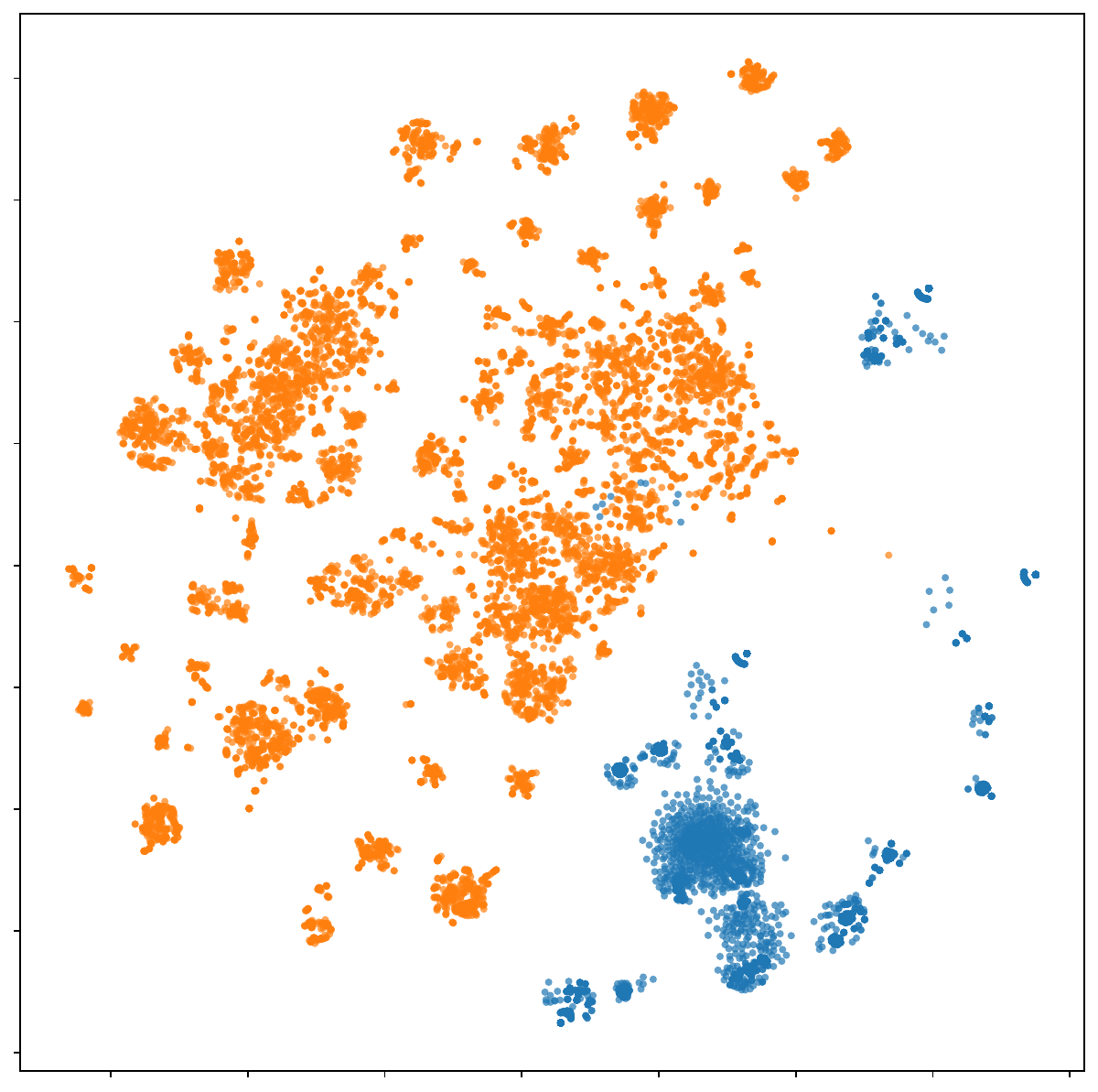}
    \caption{RAMP-PNN}
    \label{fig:pnn_dtcn}
  \end{subfigure}
  
  \begin{subfigure}[b]{0.45\linewidth}
    \centering
    \includegraphics[width=\linewidth]{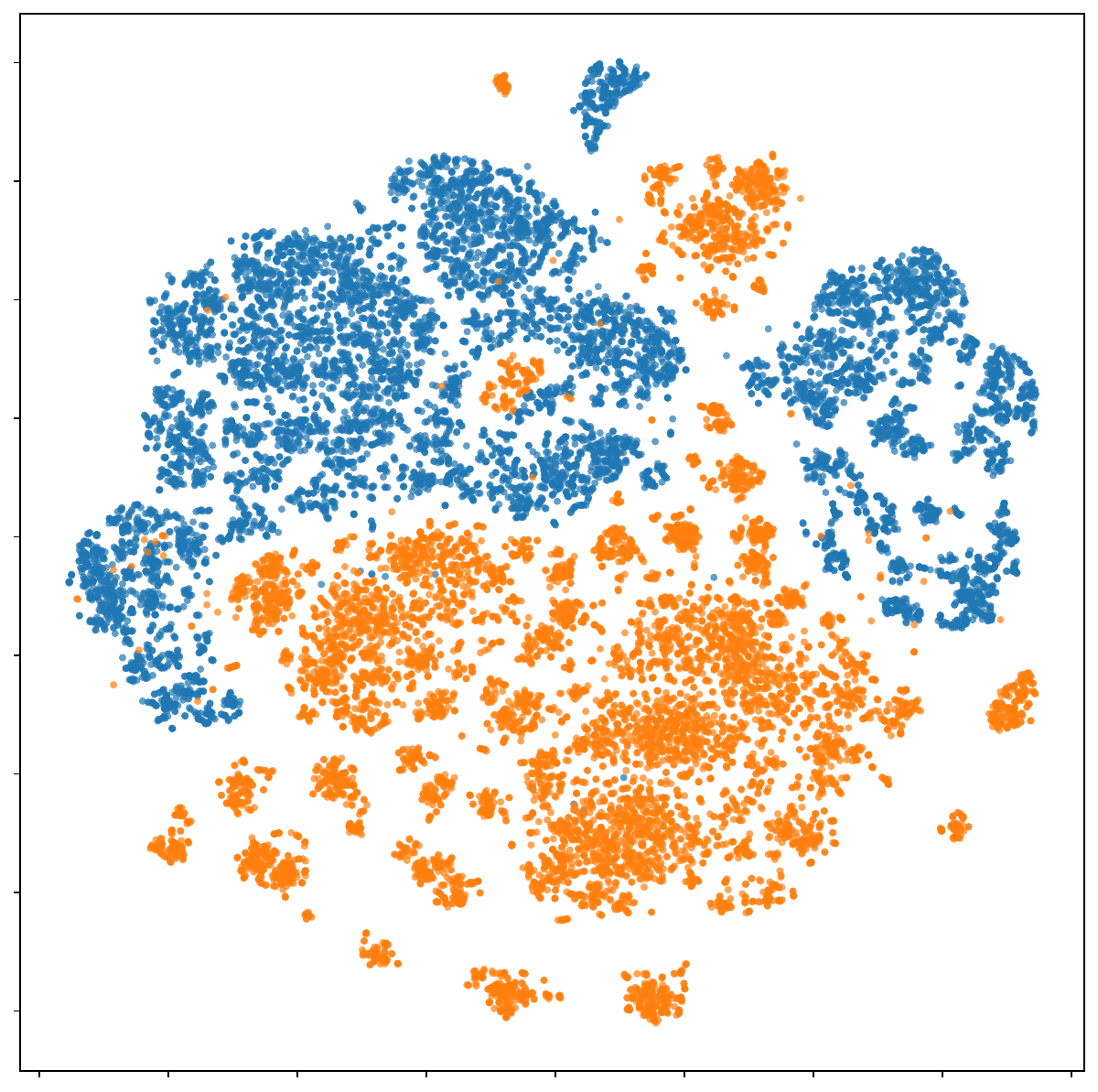}
    \caption{FCN}
    \label{fig:fcn_base}
  \end{subfigure}
  \hfill
  \begin{subfigure}[b]{0.45\linewidth}
    \centering
    \includegraphics[width=\linewidth]{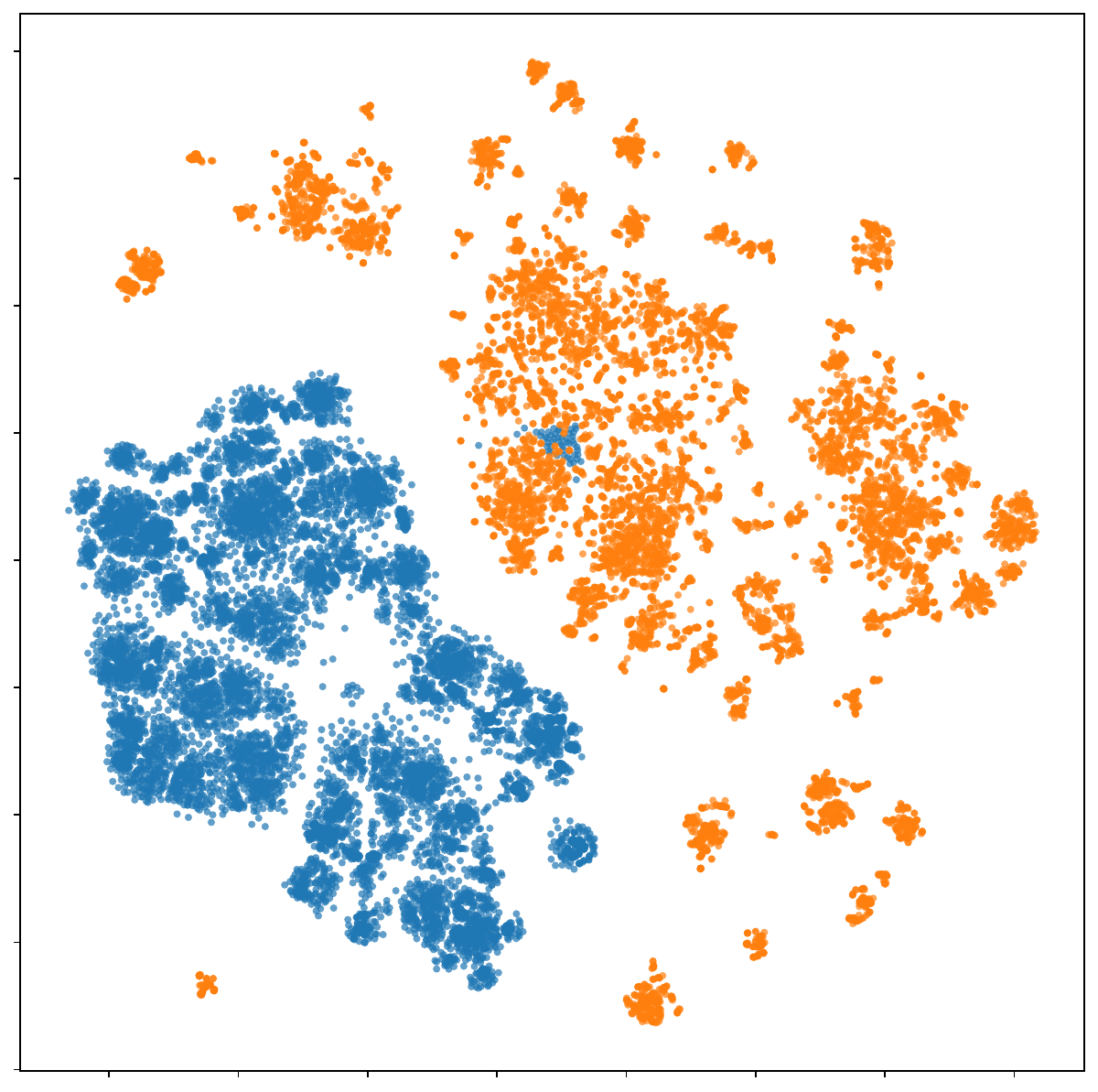}
    \caption{RAMP-FCN}
    \label{fig:fcn_dtcn}
  \end{subfigure}
  \Description{t-SNE visualization plots comparing feature embeddings for baseline models (PNN, FCN) and the proposed RAMP variants. The baseline plots show personalized and non-personalized sample embeddings heavily overlapped and mixed. In contrast, the RAMP plots show clear separation between personalized and non-personalized embeddings, forming distinct clusters.}
  \caption{\small Visualization of feature separation on CriteoPrivateAd using PCA and t-SNE with 10,000 random samples.}
  \label{fig:embedding_comparison}
\end{figure}
\subsection{Feature Representation Visualization (RQ3)}

To provide qualitative evidence of improved feature learning, we visualize the learned embeddings on CriteoPrivateAd. Specifically, we concatenate all feature embeddings from personalized and non-personalized fields separately, apply PCA to extract the top 50 principal components, and further reduce to 2 dimensions using t-SNE for visualization. Figure~\ref{fig:embedding_comparison} presents the resulting embedding spaces for baseline models versus RAMP variants. The baseline models (Figures~\ref{fig:pnn_base} and~\ref{fig:fcn_base}) exhibit substantial overlap between personalized (blue) and non-personalized (orange) feature embeddings, with the two types heavily intermingled and forming largely indistinguishable clusters. 
In contrast, RAMP variants (Figures~\ref{fig:pnn_dtcn} and~\ref{fig:fcn_dtcn}) demonstrate markedly improved feature separation, with personalized and non-personalized features forming distinct, well-separated clusters in different regions of the embedding space. Each cluster also exhibits tighter intra-class cohesion with reduced dispersion. This enhanced discriminability can be attributed to RAMP's dual mechanisms: the dual-tower component with output masking encourages each tower to specialize in its respective feature type, while the two-pathway architecture explicitly guides the non-personalized tower to learn distilled representations that are both informative and distinct from personalized knowledge. The clearer feature boundaries enable better handling of scenarios with missing personalized features, as non-personalized embeddings maintain semantic coherence independently, contributing to RAMP's superior performance in privacy-preserving settings. 

\section{Conclusion and Future Directions}\label{sec:conclusion}
In this paper, we addressed the challenge of effective advertising under modern privacy constraints by proposing a novel robust ad recommendation under limited personalized-feature availability via masking and alignment pathways (RAMP) to enhance CTR/CVR prediction for non-personalized data. Our approach integrates a personalized pathway built upon a dual-tower component, which utilizes output masking to effectively learn from both personalized and non-personalized data streams; a non-personalized pathway trained with non-personalized features; and a distillation-inspired architecture that aligns the prediction of these two pathways. Extensive experiments on public and industrial datasets demonstrate that RAMP significantly outperforms state-of-the-art baselines, particularly on non-personalized data, while maintaining strong performance in personalized scenarios. The component analysis further validates the distinct contributions of the two pathways, confirming their roles in enhancing overall model robustness and specifically boosting non-personalized prediction accuracy. Our work offers a practical and flexible solution for developing high-performance recommender systems that respect user privacy.


Beyond the demonstrated superiority of RAMP in predictive accuracy, there are still several directions for future work.
First, non-personalized features are generally closer related to item or ad attributes. It is beneficial to enhance the item side feature quality with auxiliary data sources such as structured information from the knowledge graph. Second, while we simulate non-personalized traffic through feature masking to isolate the impact of feature availability, real non-consented users may also exhibit distribution shifts. Extending RAMP to jointly address feature-availability mismatch and distribution mismatch remains an important direction for future research. Finally, as a model-agnostic framework, RAMP can be further integrated with MoE-based architectures (e.g., PLE and MMoE) and enriched item-side information sources, such as knowledge graphs, to further improve performance.



\begin{acks}
This work was supported by Science Foundation Ireland through the Insight Centre for Data Analytics (Grant No. SFI/12/RC/2289\_P2) and Huawei Ireland Research Center. We also acknowledge the computational facilities and support provided by the Irish Centre for High-End Computing (ICHEC).
\end{acks}
\newpage
\bibliographystyle{ACM-Reference-Format}
\balance
\bibliography{reference}

\clearpage
\end{document}